\documentclass[runningheads,a4paper]{llncs}
\usepackage{epstopdf}
\usepackage{amssymb}
\usepackage{graphicx}
\usepackage{amsmath}
\usepackage{url}
\usepackage{fixltx2e}
\usepackage{algorithm}
\usepackage{algpseudocode}
\usepackage{mathtools}

\DeclarePairedDelimiter\floor{\lfloor}{\rfloor}

\urldef{\mailsa}\path|arajades@asu.edu, huan.liu@asu.edu|
\newcommand{\keywords}[1]{\par\addvspace\baselineskip
\noindent\keywordname\enspace\ignorespaces#1}

\begin{document}

\mainmatter  

\title{Identifying Users with Opposing Opinions in Twitter Debates}

\titlerunning{Identifying Users with Opposing Opinions in Twitter Debates}

%
%
\author{Ashwin Rajadesingan \and Huan Liu}
\authorrunning{Ashwin Rajadesingan \and Huan Liu}

\institute{Arizona State University\\
\mailsa}

%
%

\toctitle{Lecture Notes in Computer Science}
\tocauthor{Authors' Instructions}
\maketitle

\begin{abstract}

In recent times, social media sites such as Twitter have been extensively used for debating politics and public policies. These debates span millions of tweets and numerous topics of public importance. Thus, it is imperative that this vast trove of data is tapped in order to gain insights into public opinion especially on hotly contested issues such as abortion, gun reforms etc. Thus, in our work, we aim to gauge users' stance on such topics in Twitter. We propose ReLP, a semi-supervised framework using a retweet-based label propagation algorithm coupled with a supervised classifier to identify users with differing opinions. In particular, our framework is designed such that it can be easily adopted to different domains with little human supervision while still producing excellent accuracy.

\end{abstract}

\section{Introduction}

With the advent of online social networks, almost every topic of importance is being constantly discussed and debated by millions of people online. Because of their immense reach and their ability to quickly disseminate information, social networking sites, such as Twitter, have emerged as perfect platforms for discussions and debates. With more than 800 million registered users\footnote{http://twopcharts.com/twitteractivitymonitor} of whom 232 million are active users\footnote{http://www.sec.gov/Archives/edgar/data/1418091/000119312513424260/d564001ds1a.htm}, Twitter has become the platform of choice for most discussions and debates of public interest. For example, during the first U.S presidential debate between Barack Obama and Mitt Romney on October 3, 2012, 10.3 million tweets were generated in only 90 minutes\footnote{https://blog.twitter.com/2012/dispatch-from-the-denver-debate}. 
\\
\indent This huge, high-velocity data is absolutely invaluable as it provides a deep insight into public opinion without the need for explicit surveys and polls. As inferences from social network data are made through passive observations in which users voluntarily express their opinions, this resource, in spite of the inherent selection bias involved, may provide a useful insight into public opinion as seen in existing work\cite{o2010tweets}\cite{tumasjan2010predicting}. Thus, it is imperative that algorithms and systems are built to analyze discussions and opinions expressed in Twitter.
\\
\indent In this paper, we focus on developing a framework to identify users' position on a specific topic in Twitter. The dynamic nature of content in Twitter is very different from content in conventional media such as news articles, blogs, etc., and poses new problems in identifying user opinion. Any framework catering to Twitter needs to overcome the following challenges: (i) The sheer volume and velocity of tweets generated by millions of users, (ii) the usage an ever-changing set of slang words, abbreviations, memes etc., which are not used in common parlance, and (iii) the considerably fewer word cues to identify opinion as a result of the 140 character limit on the size of the tweet. However, social networks do provide other important information such as retweets, mentions etc., which may be used in identifying users' opinion. Furthermore, the primary limiting factor in applying most supervised learning algorithms to identify opinions in Twitter is simply the lack of sufficient, reliable training data. Currently, researchers manually label or use services such as Amazon Mechanical Turk\footnote{https://www.mturk.com/mturk/} to build training data for their supervised classifiers. However, such an approach is not practically feasible considering the diversity in topics and the huge volumes of tweets we are typically interested in. To overcome this obstacle, we design our framework such that very little manual effort is required to produce sufficient training data by exploiting patterns in the users' retweeting behavior.
\\
\indent Our framework consists of two parts: semi-supervised label propagation and supervised classification. The core idea is to use a very small set of labeled seed users to produce a set of annotations using the proposed  label propagation algorithm. The resulting set of annotations will then be used for training a supervised classifier to label the other users. The label propagation algorithm ensures that sufficient data with high quality labels is produced with little manual effort. It must be noted that any supervised classifier may be used in the second step. However, as proof of concept, we use a Multinomial Na\"{\i}ve Bayes classifier trained on unigrams, bigrams and trigrams to showcase the efficiency of our framework. The main contributions of this paper are summarized below:

\begin{itemize}
\item Present \textbf{ReLP}, a semi-supervised \textbf{Re}tweet-based \textbf{L}abel \textbf{P}ropagation framework, to identify users' opinion on a topic in Twitter;
\item Drastically reduce the manual effort involved in constructing reliable training data by using users' retweet behavior;
\item Comprehensively evaluate our framework on both visibly opinionated users such as politicians, activists etc., and moderately opinionated common users who make up the majority producing excellent results.
\end{itemize}

The rest of the paper is organized as follows. In Section 2, we discuss about related research. In Section 3, we discuss our framework in detail. In Section 4, we evaluate our framework on a Twitter dataset. In Section 5, we conclude and explore further research avenues.

\section{Related Work}

In recent years, many studies on detecting opinions in Twitter have been performed using lexical, statistical and label propagation based techniques with reasonably successful results. Closely related to our work, Speriosu et al.\cite{speriosu} devised a semi-supervised label propagation algorithm using the follower graph, n-grams etc., with seeds from a variety of sources including OpinionFinder\cite{wilson}, a maximum entropy classifier etc., to identify the users' polarity on a topic. Tan et al.\cite{tan} exploit the theory of homophily in determining the sentiment of users on a particular topic by using a semi-supervised framework involving the twitter's follower and mention network. Somewhat related, Wong et al.\cite{ICWSM136105} detected the political opinion of users by formulating it as an ill-posed linear inverse problem using retweet behavior. Techniques\cite{go2009twitter}\cite{davidov2010enhanced} using emoticons and hashtags to train supervised classifiers for sentiment and opinion analysis have also been studied. Researchers have also used lexicon-based techniques using SentiWordNet\cite{esuli2006sentiwordnet}, OpinionFinder\cite{wilson} etc., to identify user opinions. 
\\
\indent Our semi-supervised framework differs from the previous approaches as it requires very little manual effort, utilizing the users' retweet behavior to obtain annotations. Unlike supervised techniques, very little seed data is required and hence, the framework may be ported to different domains with little effort. Also, our framework uses the retweet network for label propagation unlike some semi-supervised techniques which require constructing the follower graph which may not be feasible in real-time because of the volume and velocity of tweets.

\section{Methodology}

\begin{figure}[ht!]
\centering
\includegraphics[width=90mm]{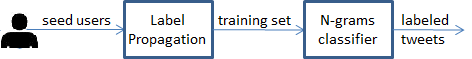}
\caption{Schematic sketch of the ReLP framework}
\label{framework}
\end{figure}

The proposed ReLP framework consists of two parts: semi-supervised label propagation and supervised learning. The seeds inputted by the user is used to produce an expanded training set which is then used to train a supervised classifier. A sketch of the framework described is shown in Fig. \ref{framework}.   

We base our approach on our observation that if many users retweet a particular pair of tweets within a reasonably short period of time, then it is highly likely that the two tweets are similar in some aspect. In the context of Twitter debates, naturally, it is highly likely that this pair of tweets share the same opinion. For example, we are more likely to find a large number of users retweeting a pair of tweets that support gun reforms than retweeting a pair, one of which is for gun reforms while the other is against gun reforms. Therefore, we model this observation by constructing a column normalized retweet co-occurrence matrix M. Let $t_1$, $t_2$...$t_n$ represent the tweets in the dataset. Then, each element M\textsubscript{i,j} of matrix M represents the fraction of users who retweeted both t\textsubscript{i} and t\textsubscript{j}.
\begin{align*}
M_{i,j}=\dfrac{number\:of\:users\:retweeting\:t_i\:and\:t_j}{total\:number\:of\:users\:retweeting\:t_j}
\end{align*}

The steps involved in label propagation are described below and the pseudocode is given in algorithm \ref{algorithm1} and algorithm \ref{algorithm2}.
\begin{enumerate}
\item Initially, we label the seed users' tweets using the label of the seeds themselves as it is extremely unlikely that seed users (who are major voices) would actually contradict their viewpoint in their own tweets. Each tweet label has values for two fields, \textit{for} and \textit{against}, which are updated during label propagation. If the tweet is in support of the topic, the \textit{for} value is higher than the \textit{against} value and vice-versa. The value of the fields may vary from 0 to 1. The seed users' tweets are initialized with the field values based on the input label of the seed users. All other tweets have both field values initialized to 0. 

\item In each iteration, we select the tweets whose labels are to be propagated using an increasing hash function h. We define the hash function $h(v)= \floor*{v*n}$ where \textit{v} $\in [0,1]$. Here, we use $n=10$, a greater precision may be obtained if \textit{n} is set lower, however, this may result in extra iterations taking more time. The value of \textit{n} can be set based on scalability requirements.

\item Using each selected tweet, we update the label of its co-occurring tweets weighted based on values of M.

\item Steps 2 and 3 are repeated until no tweets are returned by the hash function.

\end{enumerate}

Finally, the tweets are labeled as ``for'' or ``against'' depending on their \textit{for} and \textit{against} field values, whichever has higher value. While this method does not label all the tweets, it provides a very accurate classification of the tweets that it labels (as shown in Section 4). Therefore, the tweets labeled through this process are then used to train a supervised classifier to classify the remaining tweets. We use a Multinomial Na\"{\i}ve Bayes classifier with unigrams, bigrams and trigrams as features. As any out-of-the-box supervised classifier may be used, we do not go into details of the classifier's workings. We used Scikit-learn (a machine learning library in Python)'s\cite{scikit-learn} implementation of the classifier.

To elucidate the workings of the framework, we provide an example scenario: a user would like to view tweets from users in both sides of a particular debate, say gun reforms, to obtain a balanced and informed viewpoint. The user has some idea of who the main actors are, for example, Barack Obama and Piers Morgan who are for gun reforms and the National Rifle Association (NRA) and Rand Paul who are against gun reforms. The framework aims to take as input the aforementioned users and churn out tweets from other users having similarly opposing views. 

\begin{algorithm}[H]
\caption{Label propagation algorithm}
\label{algorithm1}
\begin{algorithmic}
\Procedure {LabelPropagation}{$M$}
\State $labels\leftarrow labeled\;tweets\;of\;seed\;users$
\State $final\_labels\leftarrow None$
\State $tweets \leftarrow SeedSelection(labels)$
\While {$tweets \not= None$}
\ForAll {$t_i\:\textbf{in}\:tweets$}
\ForAll {$t_j\:\textbf{in}\:M[t_i]$}
\State $labels[t_j][`for'] \leftarrow  labels[t_j][`for'] + labels[t_i][`for'] * M[t_i][t_j] $
\State $labels[t_j][`against'] \leftarrow  labels[t_j][`against'] + $
\State $\;\;\;\;\;\;\;\;\;\;\;\;\;\;\;\;\;\;\;\;\;\;\;\;\;\;\;\;\;\;\;\;\;\;\;\;\;labels[t_i][`against'] * M[t_i][t_j] $
\EndFor
\State $final\_labels[t_i]\leftarrow labels[t_i]$
\State $delete\:labels[t_i]$
\EndFor
\State $tweets \leftarrow SeedSelection(labels)$
\EndWhile
\EndProcedure
\end{algorithmic}
\end{algorithm}
\begin{algorithm}[H]
\caption{Selecting seeds for label propagation}
\label{algorithm2}
\begin{algorithmic}
\Procedure{SeedSelection}{$labels$}
\ForAll {$t_i\:\textbf{in}\:labels$}
\State $max\_value \leftarrow max(labels[t_i][`for'],labels[t_j][`against'])$
\State $store\:t_i\:using\:h(max\_value)$
\EndFor
\State \textbf{return} tweets with the highest hash value 
\EndProcedure
\end{algorithmic}
\end{algorithm}

\section{Evaluation}

In order to evaluate our framework, we use a real-world twitter dataset. We collected over 900,000 tweets\footnote{The dataset may be obtained by contacting the first author} over a period of five days during the hotly contested gun reforms debate from April 15th, 2013 to April 18th, 2013. The dataset was collected using Twitter's Streaming API\footnote{https://dev.twitter.com/docs/streaming-apis} using the keywords ``gun'' and ``\#guncontrol''. We deliberately used non-partisan keywords so as not to create any bias in the data collected. We filtered out the non-english tweets and ignored users who posted only one tweet during the entire data collection period as we found that those tweets to be very noisy (which maybe due to the keywords used in data collection). The filtered and processed dataset consists of 543,404 tweets from 116,033 users and contains 246,454 retweets.

\indent We evaluate and compare the performance of the ReLP framework on the collected dataset using three competitive baselines. The baselines were designed so as to require the same/similar amount of manual effort needed by our framework for opinion identification. Furthermore, it is important to note that the baselines (except B3) function using the same supervised classifier used in our framework. The only difference between the methods is in the way in which they are trained. Therefore, any difference in the performance of our framework with respect to the baselines can be directly attributed to the efficiency (or inefficiency) of our label propagation algorithm. We use the following three baselines:
 
\begin{itemize}
\item  \textbf{Baseline 1 (B1)}: Multinomial Na\"{\i}ve Bayes classifier with unigrams, bigrams and trigrams as features trained using the seed users' tweets. 
\item  \textbf{Baseline 2 (B2)}: Multinomial Na\"{\i}ve Bayes classifier with the same features as above, trained using partisan hashtags ( \#Protect2A, \#NewtonBetrayed).
\item  \textbf{Baseline 3 (B3)}: K-means clustering algorithm with the same features as above. The initial centroids were chosen from seed users' tweets. The final labels are assigned to the clusters after sampling instances from both clusters. 
\end{itemize}

\indent As it is practically impossible to obtain the ground truth labels for all the users, we evaluate our framework over smaller subsets of users. In order to obtain a representative sample of the users involved, we divide and sample users from two groups (inspired by Cohen et al.\cite{cohen2013classifying}): visibly opinionated users and moderately opinionated users.The rationale behind such a division is to evaluate the framework on not only obviously opinionated users such as politicians, organizations etc., but also on the relatively inactive common users who, in fact, constitute the majority.
Therefore, a good classifier must be capable of predicting the opinion of users from both brackets. Details on how the subsets of users were curated are given below.
 \begin{itemize}
 \item \textbf{Visibly Opinionated Users}: This group consists of the most vocal and opinionated users such as senators, activists, activist organizations etc. We collected a subset of these users using Twitter lists. In Twitter, any user can create lists and add other users to these lists. We manually identified lists such as ``Protect 2nd Amendment'', ``Guns Save Lives'' etc., whose users were clearly against gun reforms and other lists such as ``Prevent Gun Violence'', ``Gun Safety'' etc., whose users were clearly for gun reforms. We collected the list members and filtered out users whose tweets were not present in the dataset. In total, we obtained 263 users with 85 users for gun reforms and 178 users against gun reforms. 
  \item \textbf{Moderately Opinionated Users}: We define this group as users who posted between 2-4 tweets during the collection cycle, do not belong to any relevant lists and do not label themselves as for/against gun reforms in their Twitter profile page. We manually annotated a randomly selected sample of 500 users based on the tweets that they posted. Out of 500 users, 276 users were for gun reforms, 120 users were against gun reforms and the rest shared tweets information (such as gun reforms related news) but did not voice out their personal opinions and were hence, ignored.   
 \end{itemize}

The difference in the number of users on both sides between the two subsets can be attributed to the well studied observation of the silent majority and vocal minority\cite{mustafaraj2011vocal}. Using Barack Obama, White House and Gabrielle Giffords as seed users in support of gun reforms and the National Rifle Association (NRA), Ted Cruz and Pat Dollard as seed users against gun reforms, we obtain the results shown in Table \ref{results}. We chose these users as seeds based on the large number of retweets generated by their tweets while making sure that the total number of retweets in support of and against gun reforms are fairly balanced. We use precision, recall and f-measure to compare the various methods used.

\begin{table}[h] 
\caption{Performance evaluation and comparison with baselines} 
\label{results}
\centering 
\begin{tabular}{l   ccc    ccc} 
\hline\hline 
\\ [0.1ex] 
\textbf{ Methods}   & \multicolumn{3}{c}{  \textbf{Moderately opinionated users}  }  &\multicolumn{3}{c}{   \textbf{Visibly opinionated users}}
\\ [0.3ex] 
\hline 
 
  &\textbf{precision} &\textbf{recall}&\textbf{F-measure}&\textbf{precision} &\textbf{recall} &\textbf{F-measure} \\[1ex] 
  \\
\textbf{ReLP}&96.73 & 93.36& 95.01 & 94.18 & 97.59 & 95.85 \\[1ex] 
\textbf{B1}&81.09 & 80.79& 80.94& 75.75 & 88.23& 81.51 \\[1ex] 
\textbf{B2}&81.70 & 86.04& 83.81 & 56.10 & 97.64& 71.25 \\[1ex]
\textbf{B3}&66.26 & 78.98& 72.06 & 38.73 & 50.58 & 43.86 \\[1ex] 
 
\hline 
\end{tabular} 
\end{table} 

From Table \ref{results}, we observe that ReLP framework clearly outperforms the baselines in classifying both visibly opinionated and moderately opinionated users in almost every measure. Importantly, the f-measure (harmonic mean of the precision and recall values) of our framework is consistent across the subsets of users unlike other approaches whose f-measures fluctuate. This shows the versatility of the framework in efficiently classifying users across the spectrum.
\section{Conclusions and Future Work}
The semi-supervised ReLP framework showcases the effectiveness of combining a simple label propagation algorithm with existing supervised classifiers. The framework greatly reduces need for manually labeled training sets which are the main obstacles preventing extensive, large-scale use of supervised methods in the ever-evolving Twittersphere. This ensures that, in order to port the framework to another domain, the user only needs to supply a list of major players in that domain. Interestingly, by choosing only the major players as seeds, the user need not even understand the language of the tweets supplied as training data. 

In the future, we aim to perform experiments to determine the optimum number of seed users required by our framework to provide good classification results and examine its sensitiveness to seed selection. Furthermore, we wish to incorporate other activities in Twitter such as mentioning, replying, into our framework to improve performance. We also wish to build an automated system which will be capable of identifying and showcasing in real-time, conversations/debates between users with opposing views on a particular topic.

\subsubsection{Acknowledgements}
This research is supported by ONR (N000141110527) and (N000141410095). 
\bibliography{typeinst}
\bibliographystyle{ieeetr}

\end{document}